\documentclass[preprint,12pt]{elsarticle}

\usepackage{amssymb}
\usepackage[utf8]{inputenc}
\usepackage{hyphenat}
\usepackage[normalem]{ulem}
\biboptions{sort&compress} 
\usepackage{lineno}

\journal{Physics Letters B}

\newcommand{\PGg}{\ensuremath{\gamma}}

\newcommand{\PZ}{\ensuremath{\mathrm{Z}}}
\newcommand{\PW}{\ensuremath{\mathrm{W}}}
\newcommand{\mZ}{\ensuremath{\mathrm{m}_\mathrm{Z}}}
\newcommand{\mV}{\ensuremath{\mathrm{m}_\mathrm{V}}}
\newcommand{\mW}{\ensuremath{\mathrm{m}_\mathrm{W}}}
\newcommand{\mll}{\ensuremath{\mathrm{m}_{\ell\ell}}}
\newcommand{\mllhat}{\ensuremath{\hat{\mathrm{m}}_{\ell\ell}}}
\newcommand{\mlllo}{\ensuremath{\mathrm{m}}^{\mathrm{lo}}_{\ell\ell}}
\newcommand{\mllhi}{\ensuremath{\mathrm{m}}^{\mathrm{hi}}_{\ell\ell}}
\newcommand{\yll}{\ensuremath{y_{\ell\ell}}}
\newcommand{\absyll}{\ensuremath{|\yll|}}

\newcommand{\sstw}{\ensuremath{\sin^2\theta_W}}
\newcommand{\cstw}{\ensuremath{\cos^2\theta_W}}
\newcommand{\sqtw}{\ensuremath{\sin^4\theta_W}}
\newcommand{\cqtw}{\ensuremath{\cos^4\theta_W}}
\newcommand{\ssteff}{\ensuremath{\sin^2\theta_\mathrm{eff}^{\ell}}}
\newcommand{\sstmsbar}{\ensuremath{\sin^2\theta_{W}^{\MSbar}(\mu)}}
\newcommand{\alphamsbar}{\ensuremath{\alpha_\mathrm{EM}^{\MSbar}(\mu)}}
\newcommand{\MSbar}{\ensuremath{\overline{\mathrm{MS}}}}
\newcommand{\TeV}{\ensuremath{\mathrm{TeV}}}
\newcommand{\GeV}{\ensuremath{\mathrm{GeV}}}
\newcommand{\powheg}{\textsc{POWHEG-BOX-V2}}
\newcommand{\afb}{\ensuremath{A_\mathrm{FB}}}
\newcommand{\diff}{\ensuremath{\mathrm{d}}}
\newcommand{\pt}{\ensuremath{p_{T}}}
\newcommand{\tcs}{\ensuremath{\theta_{CS}}}
\newcommand{\xfitter}{\mathrm{xFitter}}
\newcommand{\pythia}{\textsc{pythia8}}
\newcommand{\rivet}{\textsc{Rivet}}
\newcommand{\invfb}{\ensuremath{\mathrm{fb}^{-1}}}

\usepackage{xcolor}

\newcommand{\myeq}[1]{Eq.~\ref{#1}}
\newcommand{\myfig}[1]{Fig.~\ref{#1}}
\newcommand{\mytab}[1]{Tab.~\ref{#1}}

\usepackage{amsmath}
\usepackage{booktabs}
\usepackage{multirow}

\begin{document}

\begin{frontmatter}

\title{
\vspace*{-30mm}
{\hfill \small{DESY 23-028}}\\[-1ex]
\vspace{25mm}
Probing the weak mixing angle\\ at high energies at the LHC and HL-LHC
}
\author[a]{Simone Amoroso}
\author[b]{Mauro Chiesa}
\author[b,c]{Clara Lavinia Del Pio}
\author[a,f]{Katerina Lipka}
\author[b]{Fulvio Piccinini}
\author[a]{Federico Vazzoler\corref{cor1}}\ead{federico.vazzoler@desy.de}
\author[d,e]{Alessandro Vicini}

\affiliation[a]{organization={Deutsches Elektronen-Synchrotron DESY},
                addressline={Notkestr. 85}, 
                city={22607 Hamburg},
                country={Germany}} 
\affiliation[b]{organisation={Istituto Nazionale di Fisica Nucleare INFN, Sezione di Pavia},
                addressline={via A. Bassi 6},
                city={Pavia},
                country={Italy}}
\affiliation[c]{organisation={Dipartimento di Fisica, Università di Pavia},
                addressline={via A. Bassi 6},
                city={Pavia},
                country={Italy}}

\affiliation[d]{organisation={Dipartimento di Fisica, Università degli Studi di Milano},
                addressline={via G. Celoria 16},
                city={Milano},
                country={Italy}}

\affiliation[e]{organisation={Istituto Nazionale di Fisica Nucleare INFN, Sezione di Milano}, 
                addressline={via G. Celoria 16}, 
                city={Milano},
                country={Italy}}
                
\affiliation[f]{organisation={Bergische Universität Wuppertal,}, 
                addressline={Gaußstrassse 20}, 
                city={Wuppertal},
               country={Germany}}
               
\cortext[cor1]{Corresponding author}

\begin{abstract}
Measurements of neutral current Drell-Yan production at large invariant dilepton masses can be used to test the energy scale dependence (running) of the electroweak mixing angle.
In this work, we make use of a novel implementation of the full next-to-leading order electroweak radiative corrections to the Drell-Yan process using the $\MSbar$ renormalization scheme for the electroweak mixing angle. 
The potential of future analyses using proton-proton collisions at $\sqrt{s}=13.6~\TeV$ in the Run~3 and High-Luminosity phases of the LHC is explored.
In this way, the Standard Model predictions for the $\MSbar$ running at $\TeV$ scales can be probed.
\end{abstract}


\end{frontmatter}

%% main text
\section{Introduction}
\label{section:introduction}
The electroweak mixing angle, $\theta_W$, is one of the fundamental parameters of the Standard Model (SM) of particle physics.
In the electroweak (EW) SM Lagrangian it is defined as
\begin{equation}
    \label{eq:lagrangian-definition}
    \sstw\equiv\frac{g_1^2}{g_1^2+g_2^2},%=1-\frac{\mW^2}{\mZ^2},
\end{equation}
where $g_1$ and $g_2$ are the $U(1)_Y$ and $SU(2)_L$ gauge couplings.
When considering electroweak (EW) radiative corrections, \myeq{eq:lagrangian-definition} gets modified, depending on the renormalization scheme and on the input parameter scheme used.
In particular, in the modified minimal-subtraction ($\MSbar$) renormalization scheme, the running $\sstw$, labeled in the following as $\sin^2\theta_{W}^{\MSbar}(\mu)$, is defined as~\cite{Marciano:1979yg,Marciano:1980be,Sirlin:1989uf,Marciano:1991ix,Fanchiotti:1992tu,Sirlin:2012mh,Denner:2019vbn,Workman:2022ynf}
\begin{equation}
    \label{eq:running-couplings}
    \sin^2\theta_{W}^{\MSbar}(\mu)\equiv\frac{4\pi \alphamsbar}{{g_2^2}^{\MSbar}(\mu)},
\end{equation}
where $\alphamsbar$ is the running electromagnetic coupling. 

The EW mixing angle at the scale of $\mZ$ has been measured at both lepton~\cite{ALEPH:2005ab} and hadron colliders~\cite{CDF:2018cnj,ATLAS:2015ihy,CMS:2011utm,ATLAS:2018gqq,CMS:2018ktx,LHCb:2015jyu}, with a precision at the sub-percent level.
Actually, the quantity determined with these measurements is the effective leptonic $\ssteff$, which is defined at the $\PZ$-boson mass peak through the ratio of vector and axial-vector couplings to the $\PZ$. 
It is conceptually different from $\sstmsbar$, being the former flavour dependent and based on an on-shell definition. 
Also other definitions of a running $\sstw(\mu)$ can be given (see for instance~\cite{Czarnecki:1995fw,Czarnecki:1998xc,Czarnecki:2000ic,Ferroglia:2003wa}) different from the $\MSbar$ one. 
In our study, we stick to the $\MSbar$ parameter.
Measurements of atomic parity violation, neutrino, and polarised electron scattering on fixed targets have been used to extract the EW mixing angle at lower energies~\cite{Kumar:2013yoa,Workman:2022ynf}. 
First results for large space-like scales have been obtained in Ref.~\cite{ZEUS:2016vyd,H1:2018mkk} using DIS data. 
The running at time-like scales above the $\PZ$-boson mass, however, has yet to be probed experimentally. 
The high energy regime is of particular interest since the renormalization group equation, which governs the evolution of $\sin^2\theta_{W}^{\MSbar}(\mu)$, predicts a running with a steep positive slope~\cite{Erler:2004in} at high scales, as a result of the inclusion of the $\PW$-boson contribution for scales larger than the $\PW$-boson mass.
Probing the running of the weak mixing-angle at high energies is complementary to the corresponding studies at low-energies, where a negative slope is expected, and offers a far from trivial test of the consistency of the SM.
High energies can also indirectly probe new states carrying electroweak quantum numbers. 
Indeed the effects of New Physics (NP) can be seen as a modification of the running of the electroweak gauge couplings, in a way independent of the particular decay channels~\cite{Georgi:1974yf,Marciano:1979yg,Einhorn:1981sx}.

A large sample of neutral current Drell-Yan (NCDY) events at large dilepton invariant masses ($\mll$) is expected to be produced in proton-proton collisions at the Large Hadron Collider (LHC) at a center-of-mass energy of $\sqrt{s}=13.6~\TeV$.
In this work, we investigate the sensitivity to the weak mixing angle of the NCDY process at large dilepton invariant masses. 
Several studies appeared in the literature on the potential of LHC and future hadronic machines to constrain NP models through the analysis of running couplings with DY processes~\cite{Rainwater:2007qa,Alves:2014cda,Farina:2016rws,Gross:2016ioi}~\footnote{The constraining power of DY processes for general parameterizations of NP through the Effective Field Theory approach has been explored, for instance, in Refs.~\cite{DiLuzio:2018jwd,Torre:2020aiz,Alioli:2020kez} and references therein.}. 
The existing analyses rely on leading order (LO) EW matrix elements, where the couplings are promoted to running couplings through leading logarithmic contributions to the beta functions. 
For the first time, the possibility to probe directly the running of $\sstmsbar$ is explored by means of a full EW next-to-leading order (NLO) calculation with a hybrid renormalization scheme, where the Lagrangian parameters $e$ and $\sstw$ are renormalized in the $\MSbar$ scheme and the $\PZ$-boson mass is renormalized in the on-shell scheme. 
In the large leptonic invariant mass region the presence of the Sudakov logarithms~\cite{Sudakov:1954sw,Beccaria:1998qe,Ciafaloni:1998xg,Beccaria:1999fk,Ciafaloni:2000df,Ciafaloni:2000rp} in the NLO matrix element is known to give large contributions to the cross section and could, in principle, have an impact on 
the sensitivity determination. 
The calculation has been developed and implemented in the framework of an upgraded version~\cite{Chiesa:in-prep} of the $\textsc{Z\_ew-BMNNPV}$ process~\cite{Barze:2013fru} of the $\powheg$~\cite{Nason:2004rx,Frixione:2007vw,Alioli:2010xd} Monte Carlo (MC) event generator, which is used for the present sensitivity study. 

\section{Theoretical predictions}
\label{section:simulation setup and pseudo-data generation}
We investigate the triple differential NCDY cross sections as a function of the invariant mass, $\mll$, rapidity, $\yll$, of the dilepton system, and of the cosine of the angle between the incoming and outgoing fermions in the Collins-Soper reference frame, $\tcs$~\cite{Collins:1977iv}.
At LO the triple differential NCDY cross section can be expressed as
\begin{equation}
    \begin{split}
        \frac{\diff^3\sigma}{\diff\mll\diff\yll\diff\cos\tcs}=&\frac{\pi\alpha^2}{3\mll s}\biggr((1+\cos^2\tcs)\sum_q S_q[f_q(x_1,Q^2)f_{\overline{q}}(x_2,Q^2)\\
        &+f_q(x_2,Q^2)f_{\overline{q}}(x_1,Q^2)]+\cos\tcs\sum_q A_q\mathrm{sign}(\yll)\\
        &\cdot[f_q(x_1,Q^2)f_{\overline{q}}(x_2,Q^2)-f_q(x_2,Q^2)f_{\overline{q}}(x_1,Q^2)]\biggr),
    \end{split}
    \label{eq:NCDY cross section}
\end{equation}
where $\alpha$ is the electromagnetic coupling, $\mll = \hat{s} = x_1 x_2 s$ is the partonic center-of-mass energy and $s$ is the hadronic one.
The $f_{q(\overline{q})}(x,Q^2)$ describe the momentum fraction $x$ of the parton $q(\overline{q})$ in the colliding protons, with the momentum transfer $Q^2$ given by $\mll^2$. 
The momentum fractions $x_1$ and $x_2$ are related to $\mll$ and $\yll$ as $x_{1,2} = \frac{\mll}{\sqrt{s}} e^{\pm\yll}$. 
The symmetric $S$ and anti-symmetric $A$ coupling combinations~\cite{Ball:2022qtp} are embedded into
\begin{equation}
    \begin{split}
        &S_q=e^2_\ell e^2_q+P_{\PGg\PZ}\cdot e_\ell v_\ell e_q v_q + P_{\PZ\PZ}\cdot(v^2_\ell + a^2_\ell)(v^2_q + a^2_q)\\
        &A_q=P_{\PGg\PZ}\cdot 2 e_\ell a_\ell e_q a_q + P_{\PZ\PZ}\cdot 8 v_\ell a_\ell v_q a_q,
    \end{split}
\end{equation}
expressed in terms of the electric charges $e_i$ (in units of the positron charge) and the vector (axial-vector) couplings $v_i~(a_i)$. 
The propagator factors are given by
\begin{equation}
    \begin{split}
        &P_{\PGg\PZ}(\mll)=\frac{2\mll^2(\mll^2-\mZ^2)}{\sstw\cstw[(\mll^2-\mZ^2)^2+\Gamma^2_{\PZ}\mZ^2]}\\
        &P_{\PZ\PZ}(\mll)=\frac{\mll^4}{\sqtw\cqtw[(\mll^2-\mZ^2)^2+\Gamma^2_{\PZ}\mZ^2]},
    \end{split}
\end{equation}
where $\Gamma_{\PZ}$ represents the $\PZ$ width.
At the $\PZ$ peak, the EW mixing angle has been extracted by measuring the forward-backward asymmetry, which is defined as
\begin{equation}
A_{\mathrm{FB}} = \frac{\sigma(\cos\tcs>0) - \sigma(\cos\tcs<0)}{\sigma(\cos\tcs>0) + \sigma(\cos\tcs<0)}.
\end{equation}
At high energy, however, the absolute differential cross section is a more suitable observable for the extraction of $\sstmsbar$.
This can be seen by evaluating the logarithmic derivative w.r.t. $\sstw$, i.e. the relative variation under the change of $\sstw$, of the cross section and of $\afb$ in the limit where $\mll$ is much greater than $\mZ$~\cite{Gross:2016ioi}. 
At the representative scale of $1~\TeV$, keeping the effect of finite $\mZ$, the logarithmic derivative multiplied by $\sstw$ is found to be $\sim 0.9$ for the cross-sections and $\sim 0.3$ for $\afb$.

In our study, NCDY production in proton-proton collisions at the LHC at $\sqrt{s}=13.6~\TeV$ is considered.
We assume integrated luminosities of $300$ $\invfb$ and $3000$ $\invfb$, expected at the end of the LHC Run~3 and High-Luminosity LHC (HL-LHC) phases~\cite{Dainese:2703572}, respectively.
We evaluate the triple differential NCDY cross section in six bins in $\mll$ with boundaries $116$, $150$, $200$, $300$, $500$, $1500$, $5000~\GeV$, six bins in $|\yll|$ with boundaries $0.0$, $0.4$, $0.8$, $1.2$, $1.6$, $2.0$, $2.5$, and two bins in $\cos\tcs$ for the forward and backward directions, with $72$ bins in total.
By considering the fully differential information we combine the sensitivity of the absolute cross-sections and the forward-backward asymmetry. 
Fiducial selections, usually employed in ATLAS and CMS measurements (see for example~\cite{CMS:2018ktx}), are applied to the leptons.
The leading (sub-leading) lepton must have a transverse momentum of $\pt^\ell > 40(30)~\GeV$ and an absolute pseudorapidity of $|\eta_\ell| < 2.5$. 
The leptons are defined at Born level, i.e. prior to final-state photon radiation.

We compute our theoretical predictions using a hybrid EW scheme with $(\alphamsbar,\sstmsbar,\mZ)$ as input parameters, where $\alphamsbar$ and $\sstmsbar$ are renormalized in the $\MSbar$ scheme and $\mZ$ in the on-shell one. 
In the present study, the running parameters are calculated from the corresponding $\beta$ functions at $\mathcal{O}(\alpha)$ with decoupling of the $\PW$-boson and top-quark for $\mu < \mW$ and $\mu < m_{\rm top}$, respectively. 
The $\mathcal{O}(\alpha)$ threshold correction to the running parameters for $\mu=\mW$ is set to zero as it cancels the similar discontinuity appearing at the same perturbative order in the counterterms corresponding to the electric charge and the sine of the weak-mixing angle\footnote{The higher-order corrections to the running of $\alphamsbar$ and $\sstmsbar$ described in Ref.~\cite{Erler:2004in} are also available as options in the code, as well as the threshold corrections at $\mathcal{O}(\alpha)$ and $\mathcal{O}(\alpha^2)$ for $\mu=\mW$ and $\mu=m_{t}$, respectively.}. 
This scheme has been introduced into an upgraded version \cite{Chiesa:in-prep} of the $\textsc{Z\_ew-BMNNPV}$ process~\cite{Barze:2013fru}, within the  $\powheg$~\cite{Nason:2004rx,Frixione:2007vw,Alioli:2010xd} MC event generator framework. 
Having $\sstmsbar$ as a direct input, we can generate templates consistently at LO as well as NLO EW precision level by varying the $\MSbar$\, parameter to be determined, analogously to the case of the scheme with $\ssteff$~\cite{Chiesa:2019nqb} as an input parameter. 

The MC predictions are generated at NLO plus parton shower in QCD and include NLO virtual weak corrections, excluding photonic corrections, which are separately gauge invariant. 
The parton-level events are interfaced to $\pythia.307$~\cite{Sjostrand:2014zea} to include the effect of parton showering, underlying event, hadronization and QED radiation from quarks. 
The other photonic corrections, namely QED final-state radiation from leptons (FSR) and initial-final interference (IFI), are not included in the present study\footnote{The QED FSR and IFI, together with the QED radiation from quarks, are separately gauge invariant.}. 
The largest effect of QED corrections comes from QED FSR, which however becomes negligible due to the chosen event selection of Born-level leptons. 
The remaining contributions from QED radiation from quark and IFI are small, reaching at most $1\%$ for dilepton invariant masses between $1500\,\GeV$ and $5000\,\GeV$\footnote{This has been checked with the NLO calculation implemented in the POWHEG-BOX code, where the contributions from initial and final state radiation have been separately switched on.}. 
The uncertainty associated with the inclusion of QED radiation from quarks can be conservatively calculated by squaring the magnitude of the corresponding contribution, which is deemed negligible for the present analysis.
The matrix elements are computed with factorisation and renormalization scales set to $\mu_R=\mu_F=\mll$ and convoluted with the \verb|NPDF31_nnlo_as_0118_hessian|~\cite{nnpdf31} PDF set.
The complex-mass scheme~\cite{Denner:2006ic} is used to treat the unstable $\PZ$-boson propagator in a gauge-invariant manner.

The input EW parameters for the generation of the nominal pseudo-data events are set to their $\MSbar$ values at $\mZ$. Templates are generated assuming the SM running of $\alphamsbar$, while in each $\mll$ bin the initial condition for the running of $\sstmsbar$ corresponding to the starting scale $\mu=\mllhat$ ($\mllhat$ being the center of the dilepton invariant-mass bin) are set to the expected SM value $\pm0.01$, in order to probe the sensitivity to $\sstmsbar$. 
In each $\mll$ bin, $10^9$ MC events are generated. 
The values of the EW parameters used in each $\mllhat$ bin are reported in \mytab{tab:EW parameters used in templates generation}.

In order to mimic a realistic measurement scenario, a simplified emulation of the detector response is applied through the use of parameterized lepton efficiencies and resolutions.
The identification and reconstruction efficiencies and energy-smearing functions used for the electrons and muons are inspired by those derived by ATLAS during the LHC Run~2 data taking~\cite{ATLAS:2019jvq, ATLAS:2015lne,ATLAS:2022jjr}.
Their effect on the simulated events is evaluated using $\rivet$~\cite{Buckley:2010ar}.
The efficiency in the electron channel ranges from $65\%$ to $75\%$ as a function of $\mll$ and depends only weakly on $\absyll$. 
In the muon channels, it varies between $80\%$ and $99\%$ as a function of $\absyll$ and does not depend on $\mll$.

\begin{table}[h]
\centering
\begin{tabular}{ccccc}
$\mlllo~[\GeV]$ & $\mllhi~[\GeV]$ & $\mllhat~[\GeV]$ & $(\alpha_\mathrm{EM}^\mathrm{\overline{MS}}(\mllhat))^{-1}$ & $\sin^2\theta_{W}^\mathrm{\overline{MS}}(\mllhat)$ \\
\midrule
66 & 116 & $\mZ$   & $127.951$ & $0.23122$ \\
%\midrule 
 $116$ & $150$ & $133$ & $127.838$ & $0.23323$ \\
 $150$ & $200$ & $175$  & $127.752$ & $0.23468$ \\
 $200$ & $300$ & $250$  & $127.544$ & $0.23648$ \\
 $300$ & $500$ & $400$  & $127.269$ & $0.23885$ \\
 $500$ & $1500$ & $1000$ & $126.735$ & $0.24350$ \\
 $1500$ & $5000$ & $3250$ & $126.047$ & $0.24954$ \\
 \bottomrule
\end{tabular}
\caption{The input parameters corresponding to the hybrid EW scheme used in this work.  
In each mass bin, the first three columns indicate the lower and upper bin edges and the respective center of the bin.   
The last two columns show the values of $(\alphamsbar)^{-1}$ and $\sstmsbar$ at $\mu=\mllhat$ as predicted by the SM running.}
\label{tab:EW parameters used in templates generation}
\end{table}

\section{Uncertainties}
\label{section:expected experimental uncertainties}
Several sources of experimental and theoretical uncertainties are considered in this study. 
The statistical uncertainties in the pseudo-data are derived from the predicted number of events at reconstructed level in each bin. 
The considered values of the systematic uncertainties in the lepton reconstruction and efficiencies are inspired by the ATLAS measurement of high-mass DY cross sections~\cite{ATLAS:2016gic} at $\sqrt{s}=8~\TeV$. 
These are extrapolated to the Run~3 and HL-LHC scenarios, assuming a reduction factor of two or four, respectively. 
They are propagated through the fit in a conservative way, assuming them to be uncorrelated in $\mll$, $\absyll$ and $\cos\tcs$.
The uncertainty in the luminosity determination is taken as $1.5\%$ for Run~3 and $1\%$ for HL-LHC~\cite{Dainese:2703572}. 
In addition, the theoretical uncertainties due to the knowledge of PDFs are estimated by propagating the \verb|NNPDF31_nnlo_as_0118_hessian| eigenvectors.
The PDF variations are obtained using grids generated with \verb|Madgraph_aMC@NLO| and \verb|aMCfast|~\cite{Bertone:2014zva, Alwall:2014hca}.

Missing higher-order terms typically represent an important source of systematic uncertainties for the absolute cross sections and an accurate evaluation of these uncertainties would be important in a real measurement.
As this study is only a sensitivity analysis, a simplified estimate is used.
The NCDY production cross section is known to N3LO in the strong coupling~\cite{Duhr:2021vwj} and with exact NNLO mixed QCD-EW corrections~\cite{Bonciani:2020tvf,Bonciani:2021zzf,Buccioni:2022kgy}.
We use the \verb|n3loxs| code~\cite{Baglio:2022wzu} to compute cross-sections and 7-point variations of $\mu_R$ and $\mu_F$ as a function of $\mll$ at N3LO in QCD, and we assume the results not to depend on $\absyll$ and $\cos\tcs$.
The effect of the fiducial lepton selections is not included in \verb|n3loxs|.
Since the fiducial acceptance is very large at high $\mll$ their effect can be neglected. 
The N3LO corrections to the cross section are small, of up to $2\%$.
The scale variations are found to be smaller than the expected statistical uncertainties at the HL-LHC in all but the first three $\mll$ bins.
Their envelope is conservatively propagated to the fit by assuming them to be uncorrelated in $\mll$, $\absyll$ and $\cos\tcs$.
The pure weak radiative corrections to $d\sigma/d\mll$ range from few 0.1\% around the $\PZ$ peak to $\mathcal{O}(10\%)$ at the few $\TeV$ scale, because of the Sudakov logarithms of the form $\alpha \log^2(\mll/\mV)$, $\mV$ being the mass of the gauge vector bosons. 
While a refined treatment of electroweak corrections in the high energy regime should include a proper resummation of Sudakov logarithms, as proposed, for instance, in Ref.~\cite{Bauer:2017bnh}, our simulations are based on NLO fixed order contributions.
A rough estimate of the uncertainty associated with missing EW higher-order contributions can be obtained by taking the square of the size of the NLO weak correction. 
This amounts to an uncertainty at the per cent level in the last $\mll$ bin considered, dominated by the value at the lower edge of the bin.
Within the $\MSbar$ scheme, an alternative way to quantify the uncertainty due to missing higher orders can be given by analysing the dependence of the theoretical predictions on the $\MSbar$ renormalization scale, which by default is fixed to $\mll$.
Adopting a scale variation of a factor of two, $\mu = \mll/2 $ and $\mu= 2 \mll$, the cross sections change by about $0.1\%$ at NLO, to be compared with $\mathcal{O}(\%)$ variations at LO accuracy.
The overall uncertainty of pure weak origin can be considered below the level of the other uncertainties discussed in our present analysis.

Another source of uncertainty arises from the subtraction of background processes, such as diboson production or processes involving top quarks with subsequent leptonic decays. A comprehensive assessment of uncertainties pertaining to such backgrounds has not been explicitly addressed in our study. However, we note that these uncertainties are subdominant as compared to the other uncertainties and are typically derived from control regions utilizing empirical data, thereby their relevance is diminished with increasing statistics.

\begin{figure}[ht!]
    \centering
    \includegraphics[width = \textwidth]{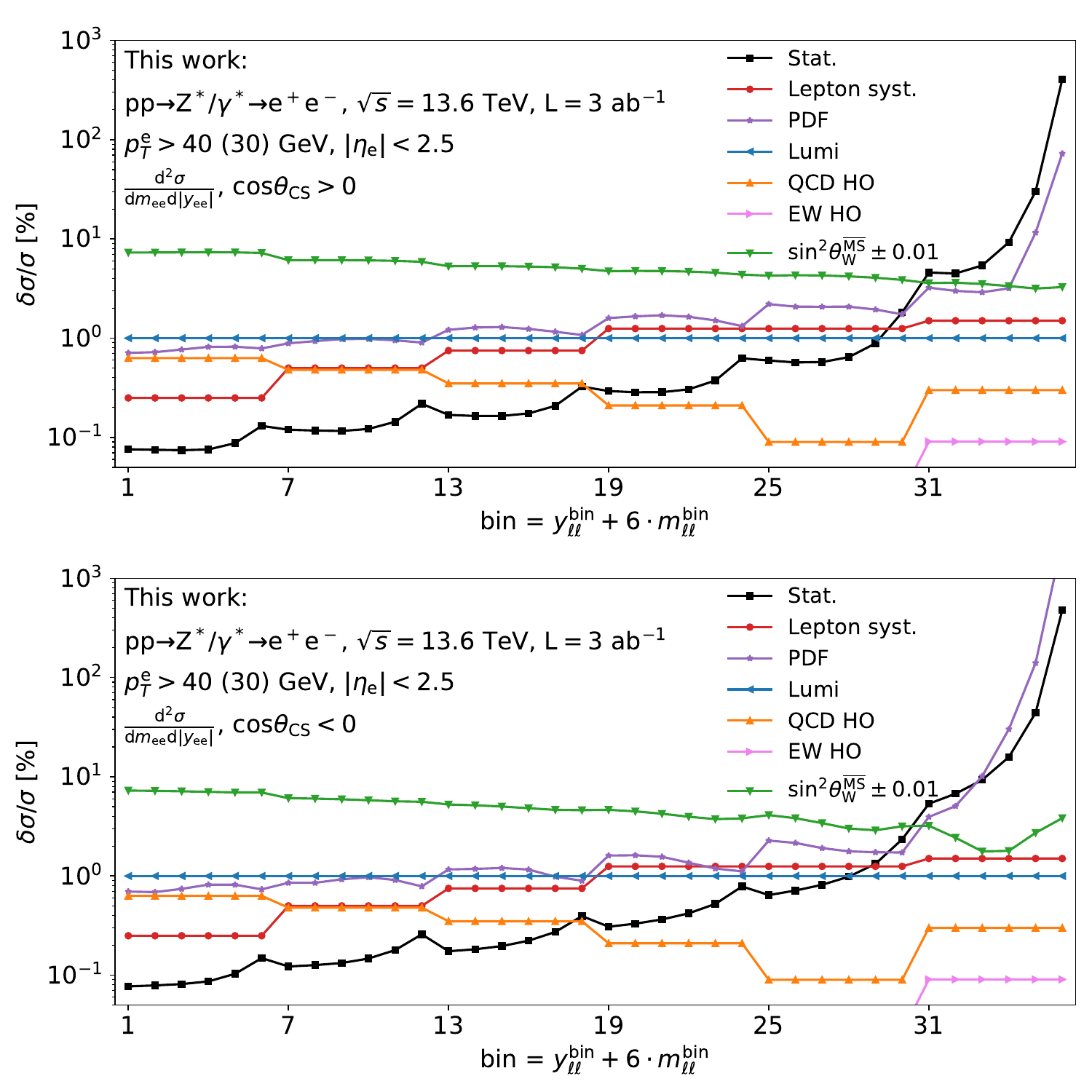}
    \caption{Relative contribution of the different sources of uncertainty to the triple differential cross section $\diff\sigma/\diff|\yll|\diff\mll$ in the forward (up) and backward (bottom) directions.
    The results are given for the electron channel in the HL-LHC scenario.
    Variations of $\sstmsbar$ by a factor $\pm0.01$ are also shown.}
    \label{fig:relative impact prefit}
\end{figure}

The contributions of different uncertainty sources to the triple differential NCDY cross sections in the electron channel are illustrated in \myfig{fig:relative impact prefit} for the HL-LHC scenario. 
A representative variation of $\sstmsbar$ by $\pm0.01$ is also shown. 
Similar results are obtained in the muon channel.

\section{Fit strategy and results}
\label{section:analysis strategy}
The sensitivity to the running is assessed by extracting the expected value of $\sstmsbar$ and evaluating its uncertainty $\delta\sstmsbar$ as a function of $\mllhat$, assuming SM running of $\alphamsbar$. 
The expected $\delta\sstmsbar$ values are obtained in a fit to the triple differential cross section pseudo-data in which independent parameters for $\delta\sstmsbar$ for each $\mll$ bin are determined simultaneously.
The fit is performed by minimising a $\chi^2$ function by using the $\xfitter$ analysis tool~\cite{Alekhin:2014irh}. 
The dependence of the cross-section on variations of $\sstmsbar$ in each bin is taken into account in the $\chi^2$ calculation using a linear approximation, that has been verified to be valid within the range of variations considered.
The expected statistical and experimental systematic uncertainties, and the theoretical uncertainties from PDFs and missing higher orders are included as nuisance parameters in the $\chi^2$ definition, such that they can be constrained in the fit.

\begin{figure}[h!]
    \centering
    \includegraphics[width = \textwidth]{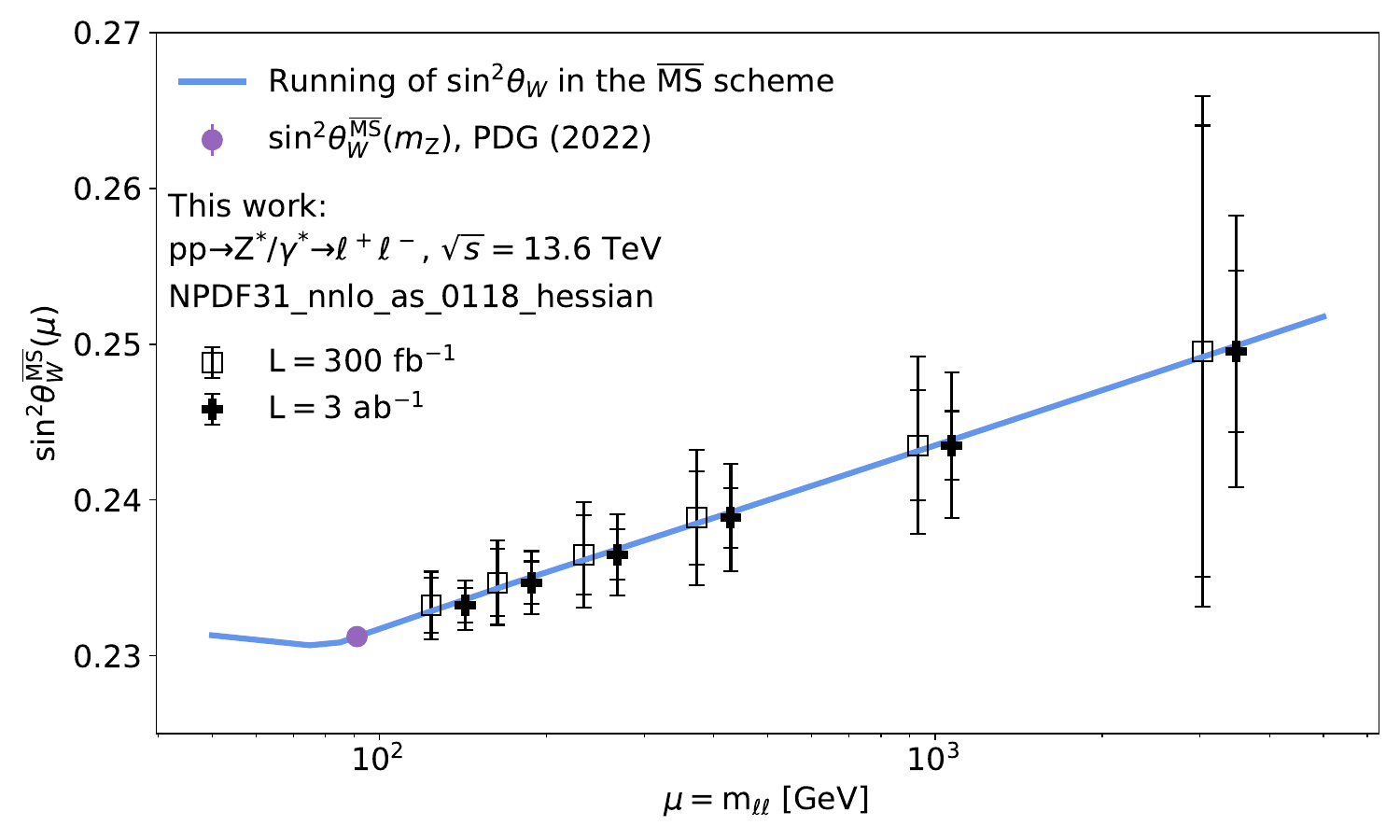}
    \caption{The scale dependence of the EW mixing angle in the SM (blue line), compared to the combined experimental measurement at $\mu=\mZ$ (violet point).
    The expected results obtained in this work are shown in black crosses (black squares) for the LHC Run~3 (HL-LHC).
    For clarity, the Run~3 and HL-LHC points are shifted to the left and right, respectively.
    The outer error bars represent the total expected uncertainty on $\sstmsbar$, while the inner error bars include only statistical and experimental uncertainties (excluding PDFs, QCD  and EW higher-order uncertainties).
    \label{fig:results for the scenarios}}
\end{figure}

The obtained values of $\delta\sstmsbar$ are presented in \myfig{fig:results for the scenarios} and in \mytab{tab:results for the scenarios}.
They range from about $1\%$ ($1\%$) to $7\%$ ($3\%$) for the LHC Run~3 (HL-LHC) scenario.
Due to the larger dataset and reduced experimental uncertainties, a significant improvement in sensitivity is expected in the HL-LHC scenario at high $\mll$.

\begin{table}[h]
\centering
\begin{tabular}{cccc|cc}
\multirow{2}{*}{$\mllhat~[\GeV]$} &  \multirow{2}{*}{$\sin^2\theta_{W}^\mathrm{\overline{MS}}(\mllhat)$} &
\multicolumn{2}{c|}{Run~3} & \multicolumn{2}{c}{HL-LHC} \\
\cline{3-6}
& & $\delta\sin^2\theta_{W}^\mathrm{\overline{MS}}(\mllhat)$ & $[\%]$ & $\delta\sin^2\theta_{W}^\mathrm{\overline{MS}}(\mllhat)$ & $[\%]$ \\ 
\midrule
 $133$  & $0.23323$ & $0.00216$ & $0.9$ & $0.00159$ & $0.7$ \\
 $175$  & $0.23468$ & $0.00271$ & $1.2$ & $0.00202$ & $0.9$ \\
 $250$  & $0.23648$ & $0.00339$ & $1.4$ & $0.00260$ & $1.1$ \\
 $400$  & $0.23885$ & $0.00434$ & $1.8$ & $0.00345$ & $1.4$ \\
 $1000$ & $0.24350$ & $0.00569$ & $2.3$ & $0.00468$ & $1.9$ \\
 $3250$ & $0.24954$ & $0.01640$ & $6.6$ & $0.00870$ & $3.5$ \\
 \bottomrule
\end{tabular}
\caption{The SM predicted value of the EW mixing angle in the $\MSbar$ renormalisation scheme $\sin^2\theta_{W}^\mathrm{\overline{MS}}(\mllhat)$ and the expected sensitivity $\delta\sin^2\theta_{W}^\mathrm{\overline{MS}}(\mllhat)$ obtained in this work, both absolute and in \%. 
The results are given as a function of the invariant mass of the final state leptons $\mllhat$, for the Run~3 and HL-LHC scenarios.}
\label{tab:results for the scenarios}
\end{table}

The largest contribution to the  uncertainty on $\delta\sstmsbar$ comes from the PDFs. 
Indeed, PDFs are known to have large uncertainties at high $x$, the kinematic range probed by high mass DY production~\cite{Ball:2022qtp}. 
To assess the dependence of our results on the choice of PDFs, the fit is repeated using the alternative PDF sets \verb|CT18ANNLO|~\cite{ct18}, \verb|MSHT20nnlo_as0118|~\cite{msht20}, \verb|ABMP16_5_nnlo|~\cite{abmp16}, and \verb|NNPDF40_nnlo_as_01180_hessian|~\cite{NNPDF:2021njg}. 
The contribution of the PDF uncertainty to $\delta\sstmsbar$, for the different PDFs, is shown in \mytab{tab:results for the pdf sets} for the HL-LHC scenario. 
It varies significantly with the PDF set used, by up to 50\% in the last $\mll$ bin.
When using sets that include a PDF for the photon, we find a comparable PDF uncertainty as we do with their non-QED counterparts.

\begin{table}[h]
\centering
\begin{tabular}{cccccc}
\multirow{2}{*}{$\mllhat~[\GeV]$} &
\multicolumn{5}{c}{$\delta\sin^2\theta_{W}^\mathrm{\overline{MS}}(\mllhat)~[\%]$} \\
\cline{2-6}
 & NNPDF31 & NNPDF40 & MSHT20 & CT18A  & ABMP16\\ 
\midrule
 $133$  & $0.5$ & $0.3$ & $0.6$ & $0.9$ & $0.5$ \\
 $175$  & $0.6$ & $0.4$ & $0.8$ & $1.0$ & $0.6$ \\
 $250$  & $0.8$ & $0.5$ & $0.9$ & $1.2$ & $0.7$ \\
 $400$  & $1.2$ & $0.6$ & $1.2$ & $1.5$ & $0.8$ \\
 $1000$ & $1.6$ & $0.8$ & $1.6$ & $1.8$ & $1.0$ \\
 $3250$ & $2.7$ & $1.6$ & $2.5$ & $2.8$ & $1.3$ \\
 \bottomrule
\end{tabular}
\caption{The contribution of the PDF uncertainty to $\delta\sin^2\theta_{W}^\mathrm{\overline{MS}}(\mllhat)$ in the HL-LHC scenario.
Results are shown in each $\mllhat$ bin for different PDF sets.}
\label{tab:results for the pdf sets}
\end{table}

\section{Conclusions}
In this work, the sensitivity of NCDY measurements at current and future LHC runs to the $\MSbar$ running of the electroweak mixing angle is investigated. 
A simulation featuring QCD and EW NLO accuracy matched to QCD parton shower is used to generate NCDY events. 
In particular, the NLO EW corrections are calculated using a hybrid EW scheme with $(\alphamsbar,\sstmsbar,\mZ)$ as input parameters, where $\alphamsbar$ and $\sstmsbar$ are renormalized in the $\MSbar$ scheme and $\mZ$ in the on-shell one. 
By using the triple differential NCDY cross section in six bins in $\mll$, six bins in $|\yll|$ and two bins in  $\cos\tcs$ for the forward and backward  directions, it is shown that measurements of Drell-Yan production in the HL-LHC phase would result in the extraction of electroweak mixing angle with a precision at the percent level, under the assumption of SM running of the electromagnetic coupling constant. 
We leave to a future study a more refined analysis, considering also additional observables. 

\section*{Acknowledgements}
The work by S. A., F. V. and K. L. is supported by the Helmholtz Association under the contract W2/W3-123.

\bibliographystyle{elsarticle-num} 
\bibliography{bibliography} 

\end{document}